\documentclass[aps,prb,twocolumn,showpacs,floats]{revtex4-1}
\usepackage[dvips]{graphicx}
\usepackage{amsmath}
\usepackage{amssymb}
\usepackage{empheq}
\newcommand{\eps}{\epsilon}

\newcommand{\ra}{\rangle}
\newcommand{\la}{\langle}
\newcommand{\ud}{\mathrm{d}}
\newcommand{\Real}{\textrm{Re}}

\newcommand{\acosh}{\textrm{acosh}}
\begin{document}

\title{\bf Rectification of spin currents in spin chains}

\author{Kevin A. van Hoogdalem and Daniel Loss}

\affiliation{Department   of  Physics,   University   of  Basel,
  Klingelbergstrasse      82,     CH-4056      Basel,     Switzerland}

\date{\today}

\begin{abstract}
We study spin transport in nonitinerant one-dimensional quantum spin chains. Motivated by possible applications in spintronics, we consider rectification effects in both ferromagnetic and antiferromagnetic  systems. We find that the crucial ingredients in designing a system that displays a nonzero rectification current are an anisotropy in the exchange interaction of the spin chain combined with an offset magnetic field. For both ferromagnetic and antiferromagnetic systems we can exploit the gap in the excitation spectrum that is created by a bulk anisotropy to obtain a measurable rectification effect at realistic magnetic fields. For antiferromagnetic systems we also find that we can achieve a similar effect by introducing a magnetic impurity, obtained by altering two neighboring bonds in the spin Hamiltonian. 
\end{abstract}

\pacs{75.10.Jm, 75.10.Pq, 75.30.Ds, 75.76.+j}
\maketitle

\section{Introduction}
Spintronics is the field of spin-based electronics, in which the spin degree of freedom is taken as a carrier of information instead of the charge, as is the case in traditional electronics. The field has seen a rapid development over the last decades.~\cite{Aws07}
Among the potential advantages of spin-based devices (which could be operated at high switching speed) in comparison to their charge-based counterparts are longer coherence times of the spins and lower energy dissipation associated with performing logical operations and transporting the information carriers.~\cite{Aws07} The latter advantage is often strongly reduced by the fact that most schemes for spin-based devices rely on spin currents that are generated as a by-product of associated charge currents.~\cite{Aws07,Zut04,Aws02}

The focus of the present work lies on transport of spin excitations in nonitinerant one-dimensional quantum systems (spin chains). In such systems magnetization can be transported by magnons (spin waves) or  spinons (domain walls), without any transport of charge. Since transport only occurs in the spin sector the energy dissipation associated with the transport of a single spin in nonitinerant systems~\cite{,Trauz08} is much lower than in itinerant systems,~\cite{Hal06,Ovc08} in which spin transport is a by-product of the transport of charge. Furthermore, nonitinerant systems have also been proposed for performing low-power logic operations.~\cite{Che09} Since energy dissipation is such an important limitation in modern spin and electronic devices, these facts make nonitinerant systems a promising candidate for applications in spintronics.~\cite{Mei03} 

Possible candidates for spin chains are, for instance, bulk structures of KCuF$_3$, SR$_2$CuO$_3$, or SrCuO$_2$, in which the exchange interaction between different chains in the crystal is much weaker than the intrachain interaction.~\cite{Ten95,Mot96,Sol01,Hes01} There exist spin chains with rather larger anisotropy in the exchange interaction, for instance, Cs$_2$CoCl$_4$ (see Ref. \onlinecite{Ken02}). In a recent experiment~\cite{Hlu10} it has been shown that the cuprate SrCuO$_2$ has a mean-free path for spinon excitations exceeding 1 $\mu$m, which shows that ballistic transport of magnetic excitations over relatively long distances in such systems is indeed possible. Various mechanisms to create pure (no charge currents) spin currents have been proposed, including spin pumping~\cite{Bra02} and the creation of spin voltage by means of the spin Seebeck effect.~\cite{Uch10} It has also been shown that, using the spin Hall effect, it is possible to convert an electric signal in a metal into a spin wave, which can then be transmitted into an insulator.~\cite{Kaj10} We mention here also the fact that this type of ballistic transport in magnetic systems is not the only kind of nearly dissipationless spin transport.  Other systems that have received wide attention recently are topological insulators,~\cite{Has10} where in particular edge states in a quantum spin Hall insulator have been predicted to be dissipationless.~\cite{Mur04}  For an overview of other possibilities, see a recent review by Sonin.~\cite{Son10}

\begin{figure}
\centering
\includegraphics[width=1.0\columnwidth]{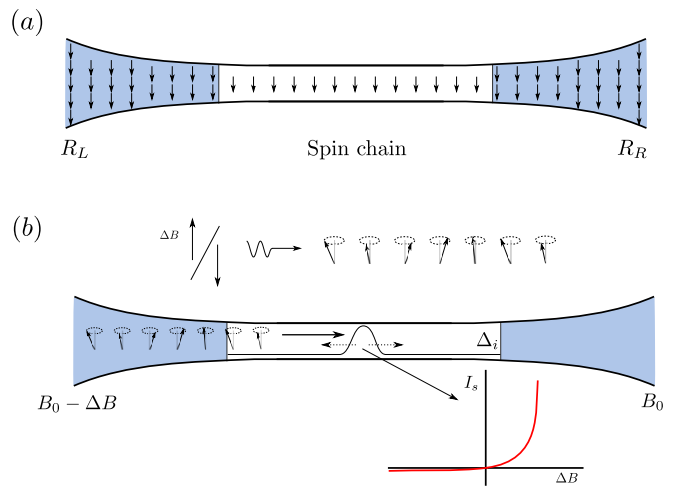}
\caption[]{(Color online) (a) Schematic view of the nonitinerant spin system. The 1D spin chain is adiabatically connected (see text) to two spin reservoirs. (b) A magnetic-field gradient $\Delta B$ gives rise to transport of magnetization through the spin chain, e.g. via spin waves. Depending on the exchange interaction $\Delta_i$ in the spin chain [see Eq. (\ref{eq:Hamil})], part of the spin current can be reflected. In a rectifying system, the magnitude of the spin current through the system depends on the sign of $\Delta B$.}
\label{fig:rect}
\end{figure}

Specifically, in this work we investigate rectification effects in nonitinerant one-dimensional quantum spin systems. Our motivation to do so lies therein that such effects form the basis for a device crucial to spin- and electronics, the transistor. In charge transport, rectification effects in mesoscopic systems have received considerable attention in recent years, both  in one-dimensional (1D)~\cite{Buet96,Feld05,Dema06,Brau07PRB} and higher dimensional systems.~\cite{San04,Spi04,And06} However, these studies have so far been limited to the itinerant electronic case. Here, we consider insulating spin chains (without itinerant charge carriers) and study the analog of such rectification effects in the transport of magnetization; see Fig. \ref{fig:rect}.

The magnetic systems  we consider are assumed to have an anisotropy $\Delta$ in the $z$ direction in the exchange interaction between neighboring spins in the spin chain. Together with an offset magnetic field applied to the entire system, also in the $z$ direction, this anisotropy will allow us to design systems displaying a nonzero rectification effect. We consider both ferro- and antiferromagnetic systems, using respectively a spin-wave and a Luttinger liquid description. For both types of magnetic order we find that an enhanced exchange interaction in the $z$ direction gives rise to the opening of a gap in the excitation spectrum of the spin chain. We can use the offset magnetic field to tune the system to just below the lower edge of the gap, so that we get the asymmetry in the spin current between positive and negative magnetic-field gradients $\Delta B$ required to achieve a nonzero rectification current. We find here that larger values of the exchange interaction $J$ or the exchange anisotropy $\Delta$ lead to a larger gap, and hence require stronger magnetic fields.

For the case of antiferromagnetic order we also discuss the case of suppressed exchange interaction in the $z$ direction. As it turns out, in this case we need an extra ingredient to achieve a sizable rectification effect. Here the extra is a "site impurity," a local change in the exchange anisotropy, which models the substitution of a single atom in the spin chain. We find that, in the presence of the offset magnetic field, the leading term in the perturbation caused by such an impurity flows - in a renormalization-group sense- to strong coupling for low energies. We can therefore, again in combination with the offset magnetic field, use it to achieve the rectification effect. In the regime where our calculations are valid, we find that the rectification current is quadratic in the strength of the impurity, and behaves as a negative power of $\left(\Delta B/J\right)$. The dependence of the rectification current on the anisotropy $\Delta$ is more complicated, and described in detail in Sec. \ref{sec:Est}.

This work is organized as follows: In Sec. \ref{sec:model} we introduce our model, in Sec. \ref{sec:SW} we discuss the rectification effect for ferromagnetic systems using the spin-wave formalism. In Sec. \ref{sec:AFLL} we map the antiferromagnetic (AF) Heisenberg Hamiltonian on the Luttinger liquid model and describe the different perturbations resulting from the anisotropic exchange anisotropy in the presence of the offset magnetic field. We then continue in Sec. \ref{sec:Ren} with a renormalization-group analysis to find the most relevant perturbations. In Sec. \ref{sec:strongUK} we focus on the case of enhanced exchange anisotropy and find the resulting rectification effect. Section \ref{sec:WC} contains the analysis for a suppressed exchange anisotropy in combination with the site impurity, and in Sec. \ref{sec:Est} we give numerical estimates of the rectification effect for the latter system.

\section{System and model}
\label{sec:model}
 The system we consider consists of a one-dimensional spin chain of length $L$, adiabatically connected (see below) to two three-dimensional spin reservoirs. 
 This system is described by the Heisenberg Hamiltonian
\begin{equation}
H = \sum_{\la ij\ra} J_{ij}\left[S_i^xS_{j}^x + S_i^yS_{j}^y + \Delta_{ij} S_i^zS_{j}^z\right] + g \mu_B \sum_i B_i S_i^z.
\label{eq:Hamil}
\end{equation}
Here $\left\la i j\right\ra$ denotes summation over nearest neighbors, and we count each bond between nearest neighbors only once. $S_i^{\alpha}$ is the $\alpha$ component of the spin operator at position $\vec{r}_i$. $J_{ij}$ denotes the exchange coupling between the two nearest neighbors at $\vec{r}_i$ and $\vec{r}_j$. The non-negative variable $\Delta_{ij}$ is the anisotropy in the exchange coupling. Regarding the ground state of the spin chain, assuming $\Delta_{ij}$ to be constant, we can distinguish two different classes of ground states in the absence of an external magnetic field, depending on the sign of the exchange coupling $J_{ij}$. For constant $J_{ij}<0$ the spin chain has a ferromagnetic ground state and we can describe its low-energy properties within the spin-wave formalism. For constant $J_{ij}>0$ the ground state is antiferromagnetic, and, in principle, we can determine the full excitation spectrum using Bethe-ansatz methods.~\cite{Bet31} However, given the complexity of the resulting solution it is convenient to use a different approach and describe the system and its low-lying excitations using inhomogeneous Luttinger liquid theory, which is what we will do in this work. 

We will study two different scenarios for the exchange interaction in the spin chain. In the first scenario
\begin{equation}
\Delta_{ij}  = \Delta > 1.
\label{eq:scen1}
\end{equation}
In general, such a constant anisotropy in the spin chain will open a gap in the excitation spectrum, which we can use to design a system that displays a nonzero rectification effect. Furthermore, it will turn out that when the constant $\Delta$ satisfies $0<\Delta < 1$ we need a spatially varying $\Delta_{ij}$ to achieve a sizable rectification effect. Therefore in the second scenario we will consider a site impurity at position $i_0$. This perturbation, in which two adjacent bonds are altered, describes the replacement of a single atom in the spin chain. It can be modeled as
\begin{equation}
\Delta_{ij} = \Delta + \Delta' \delta_{i,j-1}\left(\delta_{i,i_0}+ \delta_{i,i_0+1}\right)\quad \textrm{and} \quad \Delta < 1.
\label{eq:scen2}
\end{equation}
Here $\delta_{i,j}$ is the Kronecker delta function. 

The last term in the Hamiltonian (\ref{eq:Hamil}) is the Zeeman coupling caused by an external, possibly spatially varying,  magnetic field $B_i$. This term defines the $z$ axis as the quantization axis of the spin. In all cases that we will consider the total magnetic field $B_i$ can be split into a constant part and a spatially varying part, $B_i = B_0 + \Delta B_i$. Here the constant offset magnetic field $B_0>0$ is applied to the spin chain and the two reservoirs. $\Delta B_i$ is constant and equal to $-\Delta B$ in the left reservoir and goes to zero in the contact region between the left reservoir and the spin chain. We now define a rectifying system as a system in which the spin current $I_s$ satisfies $I_s(\Delta B) \neq - I_s(-\Delta B)$; see also Fig. \ref{fig:rect}. To quantify the amount of rectification, we use the rectification current $I_r(\Delta B)$, defined as
\begin{equation}
I_r(\Delta B) = \frac{I_s(\Delta B) + I_s(-\Delta B)}{2}.
\end{equation}

\section{Rectification in ferromagnetic spin chains: spin wave formalism}
\label{sec:SW}
\label{sec:spinwave}
\begin{figure}
\centering
\includegraphics[width=1.0\columnwidth]{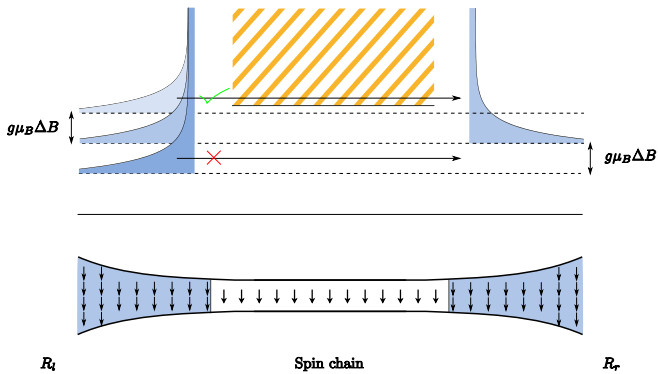}
\caption[]{(Color online) Schematic view of the system under consideration (bottom) and the band structure for the ferromagnetic exchange interaction (top). The colored areas on the left and right in the top picture depict the filled states in the reservoirs. The position of the bottom of the band in the left reservoir depends on $\Delta B$. The striped area in the middle denotes the band of allowed states in the spin chain. It is seen that spin transport through the spin chain can be large for $\Delta B < 0$, but only small for $\Delta B > 0$.}
\label{fig:system}
\end{figure}
To illustrate the mechanism behind the rectification effect we first consider a ferromagnetic spin chain with constant anisotropic exchange interaction in the $z$ direction [$J_{ij}=J<0$ and $\Delta_{ij}=\Delta > 1$ in Eq. (\ref{eq:Hamil})]. The spin chain is adiabatically connected (see below) to two three-dimensional ferromagnetic reservoirs $R_L$ and $R_R$ (see Fig. \ref{fig:rect}) and we assume initially that only the the constant magnetic field $B_0$ is present. For temperatures $T \ll g\mu B_0 /k_B$ the density of spin excitations is low, which allows us to use the Holstein-Primakoff transformation to map the Heisenberg Hamiltonian on a set of independent harmonic oscillators.~\cite{Mat81} In the presence of a constant magnetic field $B_0$ and a nonzero anisotropy $\Delta$ the excitation spectrum of the magnons, the bosonic excitations of the system, has the following form
\begin{equation}
\hbar \omega_k = g \mu_B B_0  + 2\left(\Delta-1\right) |J| s + |J| s (k a)^2,
\label{eq:magspec}
\end{equation}
with $a$ the distance between nearest-neighbor spins and $s$ the magnitude of a single spin. The magnons have magnetic moment $-g \mu_B \hat{z}$. The magnon distribution is given by the Bose-Einstein distribution $n_B(\omega_k)$. In order for the spin chain and reservoirs to be adiabatically connected, the length of the transition region $L_t$ between spin chain and reservoir must be much larger than the typical wavelength of the excitations. From Eq. (\ref{eq:magspec}) we can now see that for ferromagnets this requirement becomes $L_t \gg 2 \pi \sqrt{Js/g\mu_B B_0} a$ (we ignore the anisotropy in this estimate). For $J s = 10k_B$ J, $B_0 = 0.5$ T the requirement amounts to $L_t \gg  25 a$. 

\begin{figure}
\centering
\includegraphics[width=1.0\columnwidth]{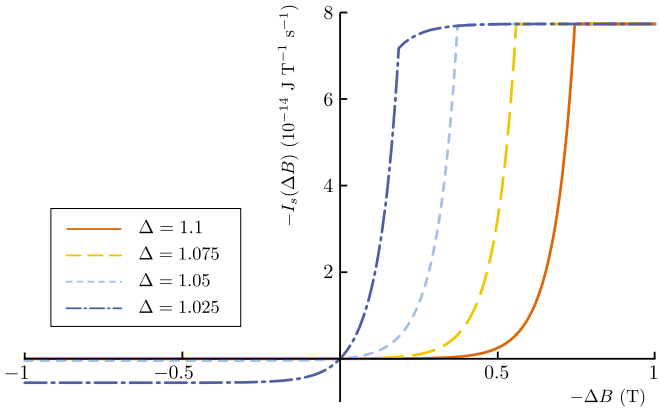}
\caption[]{(Color online) Spin current as function of applied magnetic field for the parameters $J = 10 k_B$ J, $ s = 1$, $g = 2$, $B_0 = 0.1$ $\textrm{T}$, $T = 100$ $\textrm{mK}$, and $a = 10$ $\textrm{nm}$. The spin currents for different anisotropies all saturate at the same positive value; this is the point where all the magnons incoming from the left reservoir are transmitted. This maximum current is on the order of $10^9$ magnons per second. Since the typical magnon velocity $\bar{v} = Ja^2k_0/\hbar$ is on the order of $10^3$ $\textrm{m}$ $\textrm{s}^{-1}$, and assuming length $L\approx 1$ $\mu\textrm{m}$ for the spin chain, this is indeed within the single magnon regime.}
\label{fig:RectSW}
\end{figure}

From Eq. (\ref{eq:magspec}) it follows that magnons in the spin chain require a minimum energy of $g \mu_B B_0  + 2\left(\Delta-1\right)Js$, whereas the magnons in the reservoirs (assuming that $\Delta = 1$ in the reservoirs) have a minimum energy of $g \mu_B B_0$. The effect of applying a magnetic field $-\Delta B$ to the left reservoir is to create the nonequilibrium situation in which the distribution of magnons in the left reservoir is shifted by an amount $-g \mu_B \Delta B$. From Fig. \ref{fig:system} it is seen that shifting the magnon spectrum up ($\Delta B < 0$) allows the magnons from $R_L$ to be transported through the spin chain. Because of the asymmetry of the distributions in $R_L$ and $R_R$ a large (negative) spin current will flow in this situation. When we shift the spectrum in $R_L$ down ($\Delta B > 0$), the spin current is blocked by the gap in the excitation spectrum of the spin chain, hence only a small (positive) spin current will flow. To determine the rectification current we use the Landauer-B\"uttiker approach~\cite{Dat95} to calculate the spin current through the chain. The total spin current is given by $I_s = I_{L \to R} - I_{R \to L}$, where
\begin{equation}
I_{L \to R}(\Delta B) = -g \mu_B \int_{k_{\textrm{min},L}}^{k_{\textrm{max},L}} \frac{\ud k}{2\pi} n_B(\omega_k) v(k) T(k),
\end{equation}
and $I_{R\to L}$ is defined analogously. The $\Delta B$-dependence will be shown to be in the limits of integration. Here $v(k)$ is the group velocity of the magnons, $v(k) = \partial \omega_k/\partial k$, and $T(k)$ is the transmission coefficient through the spin chain. For this system the transmission coefficient $T(k) = 1$ as long as the magnons are not blocked by the gap in the excitation spectrum of the spin chain. In the absence of $\Delta B$ the magnon spectrum in the reservoirs reaches the upper edge of the gap in the spectrum at wave vector $k_0 = \sqrt{2(\Delta-1) /a^2}$. We can incorporate the shift in the magnon spectrum in the left reservoir by changing the lower boundary in the integral for the spin current to $k_{\textrm{min},L} = \textrm{max}\left\{0,k_L\right\}$, where we defined $k_L=\Real\left[\sqrt{k_0^2+\xi\Delta B}\right]$ and $\xi \equiv g\mu_B /(Jsa^2)$. For temperatures $T$ such that $T \ll sJ/k_B$, so that we can set the upper boundary to infinity, we then have the limits of integration $\left(\textrm{max}\left\{0,k_L\right\},\infty\right)$ for the current from $R_L$ to $R_R$. For $I_{R\to L}$ we have the limits $\left( \textrm{max}\left\{k_0,k_R\right\},\infty\right)$, where $k_R = \Real\left[\sqrt{-\xi \Delta B}\right]$. The resulting spin current is then
\begin{equation}
I_s(\Delta B) = -\frac{g \mu_B}{h} \int_{\textrm{max}\left\{0,k_L\right\}}^{\textrm{max}\left\{k_0,k_R\right\}}\ud k \frac{ 2\alpha k}{e^{\beta\left(\alpha k^2 + g \mu_B B_0\right)}-1},
\end{equation}
where $\alpha \equiv sa^2J$. The spin current has been plotted in Fig. \ref{fig:RectSW} for realistic material parameters.
\section{Rectification in antiferromagnetic spin chains: Luttinger liquid formalism}
\label{sec:AFLL}
We now consider the case of an antiferromagetic spin chain consisting of spins-1/2 and adiabatically connected (for the antiferromagnetic system this means $L_t \gg 2 \pi k_F^{-1} = 4 a $, see below) to two three-dimensional antiferromagnetic reservoirs. We will show that it is possible to map both the spin chain and the two reservoirs on the Luttinger liquid model.

We start with the description of the spin chain and come back to the description of the reservoirs at the end of this section. To model the spin chain~\cite{Gia03}  we use that in one dimension we can apply the Jordan-Wigner transformation to map the spin operators onto fermionic operators: $S_i^{+} \to c_i^{\dagger}e^{i\pi \sum_{j=-\infty}^{i-1}c_j^{\dagger}c_j}$ and $S_i^z \to c_i^{\dagger}c_i-1/2$. This allows us to rewrite the part of Hamiltonian (\ref{eq:Hamil}) that describes the spin chain as the fermionic Hamiltonian:
\begin{eqnarray}
H &=& \sum_i \frac{J}{2}\left(c^{\dagger}_{i+1}c_i + c^{\dagger}_i c_{i+1}\right) + g \mu_B \sum_i B_i \left(c^{\dagger}_{i}c_{i}-1/2\right)   \nonumber\\
& & + \sum_i J \Delta_i \left(c^{\dagger}_{i+1}c_{i+1}-1/2\right) \left(c^{\dagger}_{i}c_{i}-1/2\right) \nonumber\\
& \equiv & H_0 + H_{B} + H_z.
\label{eq:fermHam}
\end{eqnarray}
We initially assume that the magnetic field satisfies $g \mu_B B \ll J$ and that $\Delta \ll 1$, so that we can use perturbation theory to describe $H_B+H_z$. We will ultimately want a description of the system by its bosonic action.

Considering first $H_0$, and restricting the Hamiltonian to low-energy excitations, we can take the continuum limit and linearize the excitation spectrum around the Fermi wave vector $k_F = \pi/(2a)$, where $a$ is again the lattice spacing, to arrive at an effective $(1+1)$-dimensional field theory involving left- and right-moving fermionic excitations. To this purpose we replace $c_i^{\dagger} \to a^{1/2}\psi^{\dagger}(x)$, $\sum_i a \to \int \ud x$ and $\Delta_i \to \Delta(x)$. Here $x=ia$. After introducing left- and right-moving fermions, $\psi^{\dagger}(x) = \psi^{\dagger}_L(x)+\psi^{\dagger}_R(x)$, we carry out a bosonization procedure~\cite{Delft08,Sene99} using the following operators:
\begin{eqnarray}
\psi^{\dagger}_r(x) &=& \frac{1}{\sqrt{2\pi a}}e^{-\eps_rk_rx}e^{i\left[\eps_r\phi(x)-\theta(x)\right]}.
\label{eq:fercrea}
\end{eqnarray}
Here $r = L,R,$ $\eps_r = \mp 1$, and $k_r$ is the Fermi wave vector for $r$-moving particles (see below). $\phi(x)$ is the bosonic field related to the density fluctuations in the system as $\partial_x\phi(x) = -\pi \left[ \rho_R(x) + \rho_L(x)\right]$. Here, $\theta(x)$ is its conjugate field, satisfying $\left[ \phi(x),\partial_{x'}\theta(x')\right] = i\pi \delta(x-x')$. We have left out the Klein operators in the fermionic creation operators, since they cancel in the subsequent perturbation theory. Using the aforementioned operations we can transform the Hamiltonian $H_0$ into the quadratic action
\begin{eqnarray}
S_0[\phi] &=& \frac{\hbar}{2\pi K} \int \ud^2 r \left[ u \left[\partial_x \phi(\vec{r})\right]^2-\frac{1}{u}\left[\partial_t \phi(\vec{r})\right]^2\right].
\label{eq:QuadHam}
\end{eqnarray}
Here and in the following we use the notation $\vec{r} = \left(x,t\right)^T$. For the free action $u = J a /\hbar$ and $K =1$, these parameters will change due to the presence of the $H_B$ and $H_z$ terms. 

The effect of the inclusion of the magnetic-field term $H_B$ is twofold: In the absence of any backscattering and umklapp scattering the field will introduce different densities of left- and right-moving excitations in the spin chain. The effect of this is to change the Fermi wave vectors in Eq. (\ref{eq:fercrea}) for the respective particles to $k_R = \pi/(2a) + \frac{K}{u \hbar} g \mu_B \left( B_0 - \Delta B\right)$ and $k_L = \pi/(2a) + \frac{K}{u \hbar} g \mu_B B_0$. This does not affect the bosonized form of $H_0$, but will have an effect on the bosonized form of $H_z$. Furthermore, the spatial dependence of the magnetic field makes that the spin chain is now described by $S[\phi] = S_0[\phi] + S_{B}[\phi]$, where:
\begin{equation}
S_{B}[\phi]  =  - \frac{g \mu_B }{\pi} \int \ud^2 r \phi(\vec{r}) \partial_x B(\vec{r}).
\end{equation}
We use the specific form of the magnetic field
\begin{equation}
\partial_x B(\vec{r}) = \Delta B\delta(x-L/2),
\label{eq:mag}
\end{equation}
which corresponds to the magnetic field described in Sec. \ref{sec:model}.

In the next step we derive the bosonic representation of the $H_z$ term.~\cite{Egg92,Ori98} In the continuum limit the $z$ component of the spin operator is given by the normal ordered expression
\begin{eqnarray}
S^z(x) &=& :\psi^{\dagger}_R(x)\psi_R(x) + \psi^{\dagger}_L(x)\psi_L(x) \nonumber\\
&& + \psi^{\dagger}_R(x)\psi_L(x) + \psi^{\dagger}_L(x)\psi_R(x):.
\end{eqnarray}
The interaction term $H_z$ contains terms that transfer approximately $0, 2k_F$ or $4k_F$ momentum. Around half filling and for a constant exchange anisotropy, e.g., the case in Eq. (\ref{eq:scen1}), conservation of momentum requires that the only terms that can survive are the ones that transfer approximately $0$ or $4k_F$ momentum, the $2k_F$ terms are suppressed by rapidly oscillating $\exp(\pm2ik_F x)$ terms. The terms that transfer small momentum give rise to terms proportional to $\left(\partial_x \phi(\vec{r})\right)^2$, and hence change the parameters $u,K$ in Eq. (\ref{eq:QuadHam}). It is not possible to determine the exact values from the linearized theory, they can, however, be determined from the Bethe-ansatz solution.~\cite{Hal80} Important here is that for $\Delta \lessgtr 1$ we have $K\gtrless 1/2$.
The $4k_F$ term (umklapp scattering term) becomes
\begin{eqnarray}
S_{\textrm{US}}[\phi] & = & \frac{aJ}{(2\pi a)^2} \int \ud^2 r \Delta(x) \cos \left[4\phi(\vec{r})-4\rho x\right],
\label{eq:UK}
\end{eqnarray}
where $\rho \equiv \frac{K}{u \hbar} g \mu_B \left(B_0 - \Delta B/2\right)$. We have neglected a constant term $\propto \rho a$ inside the cosine.

If a spatially varying anisotropy is present, as in Eq. (\ref{eq:scen2}), the $2k_F$-backscattering terms do not necessarily vanish in regions where $\Delta(x)$ varies on a scale of order $a$. The bosonization of the $2k_F$ terms then requires some care. If we naively use the continuum form of Eq. (\ref{eq:fermHam}), this term contains infinities after the bosonization. Therefore it has to be normal ordered,~\cite{Delft08} which we can do using Wick's theorem and the contraction $\overbracket{\psi_r(x) \psi}{}^{\dagger}_{r'}(x+a)  = -r\frac{e^{-i r k_r a}}{2\pi a i}\delta_{r,r'}$. Since the $S_{i}^zS_{i+1}^z$ term contains four-fermion operators the normal ordering will yield not only additional constants, but also two fermion operators, as can, for instance, be seen from the typical term $\psi_R^{\dagger}(x)\psi_R(x)\psi_R^{\dagger}(x+a)\psi_L(x+a)$, which becomes
\begin{eqnarray}
 :\psi_R^{\dagger}(x)\psi_R(x)\psi_R^{\dagger}(x+a)\psi_L(x+a): \nonumber\\
+ :\psi^{\dagger}_R(x)\psi_L(x+a): \overbracket{\psi_R(x) \psi}{}^{\dagger}_{R}(x+a).
\end{eqnarray}
The latter term is equal to $-\frac{e^{-ik_Ra}}{2\pi a i}:\psi_R^{\dagger}(x)\psi_L(x+a):$. There are four such terms; together they give the backscattering term
\begin{eqnarray}
S_{\textrm{BS}} [\phi] &=& -\frac{2aJ}{(2\pi a)^2} \int \ud^2r (-1)^{x/a} \Delta(x)\nonumber\\
 & & \times \left\{ \sin\left[ 2\phi(\vec{r}) -2 \rho x \right] +\sin\left[ 2\phi(\vec{r}) -2 \rho (x+a) \right] \right\}.
\end{eqnarray}
The completely normal-ordered term becomes
\begin{eqnarray}
S_{\textrm{BSN}}[\phi] &=& \frac{8aJ}{(2\pi a)^2}\int \ud^2 r (-1)^{x/a} \Delta(x)\left[a \partial_x\phi(\vec{r})\right]^2\nonumber\\
& & \times \left\{ \sin\left[ 2\phi(\vec{r}) -2 \rho x \right] +\sin\left[ 2\phi(\vec{r}) -2 \rho (x+a) \right] \right\}.
\end{eqnarray}
The latter can be seen by expanding the normal ordered term around $x$ and invoking the Pauli exclusion principle.
From simple dimension counting it is clear that the dimension of this operator is $2+K$. Hence it is always irrelevant, and we will neglect it in the following.
The total action describing the spin chain is then given by
\begin{equation}
S[\phi] = S_0[\phi] + S_{B}[\phi] + S_{\textrm{US}}[\phi] + S_{\textrm{BS}}[\phi].
\end{equation}

Now we can distinguish between the two scenarios in Eqs. (\ref{eq:scen1}) and (\ref{eq:scen2}). For the constant anisotropy in Eq. (\ref{eq:scen1}) the $S_{\textrm{BS}}[\phi]$ and $S_{\textrm{BSN}}[\phi]$ terms vanish because they are proportional to the rapidly oscillating $(-1)^{x/a}$. Hence only the bulk umklapp scattering term, given by
\begin{equation}
S_{\textrm{BUS}}[\phi] = \frac{\lambda_1}{(2\pi a)^2}\int \ud x \ud t \cos\left[4\phi(\vec{r})-4\rho x\right],
\end{equation}
remains. Here $\lambda_1 \equiv a J \Delta$. The spin chain in Eq. (\ref{eq:scen1}) is thus described by the action $S_0[\phi] + S_{\textrm{BUS}}[\phi] + S_B[\phi]$. As we will discuss in the next section, the bulk umklapp scattering term renormalizes as $2-4K$, hence it is irrelevant for $K>1/2$ ($\Delta < 1$) and relevant for $K<1/2$ ($\Delta > 1$). If this term is relevant, as it is for the parameters in Eq. (\ref{eq:scen1}), it opens a gap in the excitation spectrum of the spin chain, which, as we will show in Sec. \ref{sec:strongUK}, can be used to achieve the rectification effect in a similar way as for the ferromagnetic system in Sec. \ref{sec:spinwave}. To wit, if we tune $B_0$ such that it lies just below the upper edge of the gap, for $\Delta B > 0$ there can be no spin transport, since there are no states available for transport in the chain, whereas for $\Delta B < 0$ the states above the edge of the gap are accessible, and transport is made possible. 

Next we discuss the case with spatially varying exchange anisotropy, Eq. (\ref{eq:scen2}). We start out by noting that, since the bulk umklapp scattering term is irrelevant for this scenario, the spin chain is not in a gapped state in equilibrium. The effect of the applied magnetic field is then to move the spin chain away from half filling. As we will show later, in the current scenario this doping is required in order to achieve a nonzero rectification current. In the case of Eq. (\ref{eq:scen2}) the backscattering term vanishes everywhere, except in the region where $\Delta(x)$ itself varies on a short lengthscale. For the specific form of the anisotropy of Eq. (\ref{eq:scen2}), the action resulting from the site impurity is
\begin{eqnarray}
S_{\textrm{BS}} [\phi] &=& -\frac{a2J\Delta'}{(2\pi )^2} \int \ud t \nonumber \partial_x \{ \sin\left[ 2\phi(x_0,t) -2 \rho (x_0-\frac{a}{2}) \right] \nonumber\\
& & +\sin\left[ 2\phi(x_0,t) -2 \rho (x_0+\frac{a}{2}) \right] \},
\end{eqnarray}
where, here and elsewhere, $\partial_x f(x_0,t)$ should be read as $\partial_x f(x,t)|_{x=x_0}$. Adding to this term the local umklapp scattering term caused by the site impurity, the total action coming from this impurity becomes
\begin{widetext}
\begin{eqnarray}
S_{\textrm{I}}[\phi] & = & \frac{1}{\pi^2a} \int \ud t \bigg\{\frac{1}{4}\left[\lambda_{2}^a\cos4\phi(x_0,t) + \lambda_{2}^b\sin4\phi(x_0,t)\right] + \sigma\left[ \lambda_{3}^a\cos 2\phi(x_0,t) + \lambda_{3}^b\sin2\phi(x_0,t)\right]  \nonumber\\
& & + a \left[\lambda_{4}^a \partial_x\phi(x_0,t)\cos 2\phi(x_0,t) + \lambda_{4}^b \partial_x\phi(x_0,t)\sin 2\phi(x_0,t)\right]\bigg\},
\label{eq:SI}
\end{eqnarray}
\end{widetext}
which contains terms caused by umklapp scattering (proportional to $\lambda_2^{a,b}$), terms that may be called offset-induced backscattering terms (proportional to $\lambda_3^{a,b}$), and terms that describe combined forward- and backscattering (proportional to $\lambda_4^{a,b}$). In the expression for the action $S_{\textrm{I}}[\phi]$ we have defined $\sigma \equiv \rho a = \frac{K a}{u \hbar} g \mu_B \left(B_0 - \Delta B/2\right) = K \left(\frac{g\mu_B (B_0-\Delta B/2)}{\hbar \omega_c}\right)$. Here we have identified the UV cutoff of the theory $\omega_c$ with $u/a$. Furthermore, the prefactors are given by
\begin{equation}
\begin{array}{rcl}
\lambda_2^a & = &  2\lambda\left[\cos 4 \rho \left(x_0 - \frac{a}{2}\right)+\cos 4 \rho \left(x_0 +\frac{a}{2}\right)\right],\\
\lambda_2^b & = &  2\lambda\left[\sin 4 \rho \left(x_0 - \frac{a}{2}\right)+\sin 4 \rho \left(x_0 +\frac{a}{2}\right)\right],\\
\lambda_3^a &=&  \lambda(-1)^{x_0/a}\left[\cos 2\rho \left(x_0-\frac{a}{2}\right) + \cos 2\rho \left(x_0+\frac{a}{2}\right)\right], \\
\lambda_3^b &=&  \lambda(-1)^{x_0/a}\left[\sin 2\rho\left( x_0-\frac{a}{2}\right) + \sin 2\rho\left(x_0+\frac{a}{2}\right)\right].
\end{array}
\label{eq:prefac}
\end{equation}
Where $\lambda=aJ\Delta'$. Furthermore, $\lambda_4^{a,b} = - \lambda_3^{a,b}$. In this scenario the spin chain is thus described by $S_0[\phi] + S_{\textrm{BUS}}[\phi] + S_{\textrm{I}}[\phi] + S_B[\phi]$. 

Finally we return to the description of the spin reservoirs. As we show in detail in Appendix \ref{ch:A0}, we can describe the low-energy excitations of the reservoirs by the quadratic Luttinger liquid action, Eq. (\ref{eq:QuadHam}). The effective Luttinger liquid parameters $u_r, K_r$ of the three-dimensional reservoirs can be determined by mapping its dynamic susceptibility onto that of a Luttinger liquid, using the nonlinear sigma model, resulting in
\begin{equation}
u_r = \sqrt{3}Ja/\hbar \qquad K_r = \pi/(4\sqrt{3}).
\end{equation}
This means that we can describe the reservoirs by letting $u \to u(x)$ and $K \to K(x)$ in Eq. (\ref{eq:QuadHam}), where 
\begin{equation}
u(x),K(x) = \left\{\begin{array}{cl} u,K & \textrm{for } x\in (-L/2,L/2), \\ u_r,K_r & \textrm{for } x\notin (-L/2,L/2). \end{array}\right. 
\end{equation}

\section{Renormalization group treatment}
\label{sec:Ren}
We start the analysis of the antiferrromagnetic spin chain by studying the scaling behavior of the spin chain, allowing us to determine which perturbations will be most relevant in the low-energy sector. We perform the renormalization-group (RG) calculation in momentum space,~\cite{Gia03} assuming there is a hard natural momentum cutoff $\Lambda_0$ in the system. In the RG procedure the cutoff $\Lambda(l) = \Lambda_0 e^{-l}$ is decreased from $\Lambda(l)$ to $\Lambda(l+\ud l)$. For the RG procedure we consider the partition function written in terms of the action in imaginary time. As is customary we split the field $\phi(\vec{r})$ contained in this action up in a fast and a slow part, $\phi(\vec{r}) = \phi^>(\vec{r}) + \phi^<(\vec{r})$, where the fast part contains Fourier components with momentum between $\Lambda(l+\ud l)$ and $\Lambda(l)$, and the slow part contains the components with momentum less than $\Lambda(l+\ud l)$. The RG procedure then consists of integrating out the fast modes, and subsequently restoring the original cutoff in the action, in order to find an effective low-energy action with the same couplings, but different coupling constants. This allows us to find the relevant (increasing in magnitude under a decrease of the cutoff) and irrelevant (decreasing in magnitude under a decrease of the cutoff) couplings. For completeness, we mention the renormalization equations for the constant anisotropy $\Delta$ (see Ref. \onlinecite{Gia03})

\begin{equation}
\begin{split}
\frac{\ud K}{\ud l} & =  -\frac{8 \lambda_1^2}{(2\pi \hbar)^2}\frac{1}{u^2} K C_{K},\\
\frac{\ud u}{\ud l} & =  \frac{8\lambda_1^2K^2}{(2\pi)^3\hbar^2}\frac{1}{u} C_u,\\
\frac{\ud \rho}{\ud l} & =  \rho, \\
\frac{\ud \lambda_{1}}{\ud l} & =  \left(2-4K\right)\lambda_1.
\end{split}
\end{equation}
We have omitted several terms $\Lambda a$, which are a number of order 1. The different constants are given by [here $\vec{\bar{r}} = \Lambda (x, u\tau)^T$ is dimensionless]
\begin{equation}
\begin{split}
C_K & =  \frac{1}{2} \int \ud^2 \bar{r} \cos \left[ 4\rho x\right] \bar{r}^2 J_0(\bar{r})e^{-8 K F_1(\bar{r})},\\
C_u & =  -\frac{1}{2} \int \ud^2 \bar{r} (\bar{x}^2-u^2\bar{\tau}^2)\cos\left[4\rho x\right] J_0(\bar{r})e^{-8KF_1(\bar{r})}.
\end{split}
\end{equation}
Here $J_0(\bar{r})$ is the zeroth order Bessel function and $F_1(\bar{r}) = \int_0^1 \ud q \frac{1-J_0(q\bar{r})}{q}$. Both integrals converge and are of order 1. From the last RG equation it follows that for $K < 1/2$, which as we have seen before corresponds to $\Delta > 1$, the perturbation caused by the bulk umklapp scattering grows under renormalization. This corresponds to the opening of a gap in the excitation spectrum of the spin chain. The magnitude $M$ of this gap has been calculated analytically using Bethe-ansatz methods,~\cite{Clo66} and is given by
\begin{equation}
\frac{M}{J} = \frac{\pi \sinh \nu}{\nu}\sum_{n=-\infty}^{\infty} \frac{1}{\cosh\left[(2n+1)\pi^2/2\nu\right]},	
\end{equation}
where $\nu = \acosh \Delta$. For $\Delta \gtrsim 1$ this gap is exponentially small, $M \approx 4\pi J \exp\left[-\pi^2/\left((2\left[2(\Delta-1)\right]^{1/2}\right)\right]$, and for $\Delta \to \infty$ it becomes $M \approx J\left[\Delta-2\right]$.

If we now assume to be in the regime where $K>1/2$ and add the site impurity described by $S_{\textrm{I}}[\phi]$, we get the following additional equations:
\begin{equation}
\begin{split}
\frac{\ud \lambda_{2}^{a/b}}{\ud l} &= \left(1-4K\right) \lambda_{2}^{a/b}-\Gamma_{44}^{a/b}C_1-\Gamma_{33}^{a/b}C_{\beta},\\
\frac{\ud \lambda_{3}^{a/b}}{\ud l} &= (1-K) \lambda_{3}^{a/b}  - \Gamma_{23}
^{a/b}C_{\alpha} -\Lambda_{3\mp}^{b/a}C_c\pm\Lambda_{4\pm}^{b/a}C_s,\\
\frac{\ud \lambda_{4}^{a/b}}{\ud l} &= (1-K) \lambda_{4}^{a/b}   - \Gamma_{24}
^{a/b}C_{\alpha}  -  \Lambda_{4\pm}^{b/a}C_c,
\end{split}
\end{equation}
where we have defined the second-order terms $\Gamma_{nm}^{a/b} = \lambda_n^a\lambda_m^{a/b} + \lambda_n^b\lambda_m^{b/a}$ and  $\Lambda_{n\eta}^{a/b} = \left(\eta\lambda_n^{a/b}\sin4\rho x_0 + \lambda_n^{b/a}\cos4\rho x_0\right)\lambda_1$. Here $\eta = \pm 1$. The different constants used here are given by
\begin{equation}
\begin{split}
C_1 & = \frac{1}{2\pi^3\hbar u}\int \ud \bar{\tau} \frac{J_1(\bar{\tau})}{\bar{\tau}}e^{-2KF_1(\bar{\tau})},\\
C_{\alpha} &= \frac{K }{\pi^2 \hbar u } \int \ud \bar{\tau} J_0(\bar{\tau}) e^{-3KF_1(\bar{\tau})},\\
C_{\beta} & = \frac{8K\sigma^2}{\pi^2\hbar u} \int \ud^2 \bar{r} J_0(\bar{r})e^{-2KF_1(\bar{r})},\\
C_c & = \frac{K}{2\pi^2\hbar u}\int \ud^2\bar{r}\cos\left[4(\rho/\Lambda)\bar{x}\right]J_0(\bar{r})e^{-3KF_1(\bar{r})},\\
C_s & = \frac{K^2}{\sigma\pi^2\hbar u}\int \ud^2 \bar{r}\frac{\bar{x}}{\bar{r}}\sin\left[4(\rho/\Lambda)\bar{x}\right]J_0(\bar{r})e^{-3KF_1(\bar{r})}.
\end{split}
\end{equation}
Again, all integrals converge and are of order 1. From the form of the equations it follows that we can ignore contributions that are of second order in the couplings, and we see that the most relevant couplings are the $\lambda_3^{a,b}$ and $\lambda_4^{a,b}$ terms, which to first order scale as $1-K$. For $K\in(1/2,1)$, the regime in which we operate in the case of Eq. (\ref{eq:scen2}), these terms thus grow in magnitude under a decrease of the cutoff. Because of the extra $\partial_x \phi(x_0,t)$ proportionality in the $\lambda_4^{a,b}$ terms, one would expect the effect of these terms on the spin current to be smaller than that of the $\lambda_3^{a,b}$ terms. However, since the $\lambda_3^{a,b}$ terms have an additional $\sigma$ in front of them, in the regime $B_0 \approx \Delta B/2$ these are suppressed, so that they could become comparable to the $\lambda_4^{a,b}$ terms. Therefore we will consider the effect of both types of coupling when calculating the spin current for the system with anisotropy in the exchange interaction as in Eq. (\ref{eq:scen2}) in Sec. \ref{sec:wc}.
\section{Enhanced anisotropy}
\label{sec:strongUK}
As we have shown in Sec. \ref{sec:Ren}, the enhanced exchange anisotropy of Eq. (\ref{eq:scen1}) gives rise to a bulk umklapp scattering term in the action of the spin chain, and the chain is described by the sine-Gordon model.~\cite{Gia03} This model has been well studied, and is known to give rise to two possible phases depending on the value of the chemical potential $g\mu_B B$: For values of the chemical potential lower than the gap $M$ the system is a Mott insulator; when the chemical potential is increased to values above this gap, the system undergoes a commensurate-incommensurate (C-IC) transition and becomes conducting.

To simplify the sine-Gordon model we replace $\phi_M(\vec{r}) = 2\phi(\vec{r})$ and, in order to keep the commutation relation between the two fields unchanged, $\theta_M(\vec{r}) = \theta(\vec{r})/2$. The umklapp scattering term then reduces to a backscattering term (a two-particle operator), and the effective Luttinger parameter becomes $4K$. The fermionic form of the resulting new action is known as the massive Thirring model. At the Luther-Emery point, $K=1/4$, (Ref. \onlinecite{Lut74}) the new fermions whose action is given by the massive Thirring model are noninteracting, and we can diagonalize the quadratic part of the action by a Bogoliubov transformation, which gives rise to two separate bands of fermionic excitations with dispersion $\eps_{\pm}(k) = \pm \sqrt{(uk)^2+M^2}$. If $K \neq 1/4$ there are residual four-fermion interactions present. However, it can be shown~\cite{Schu80} that, independent of the initial interactions, near the C-IC transition the strength of these interactions vanishes faster than the density of the fermions, so that the interactions become negligible. The new free fermions are not the original fermions; instead they correspond to solitonic excitations of the original action. These solitons have fractional magnetic moment $-g \mu_B/2$. We finally can relinearize the excitation spectrum around the point $M$ to arrive at a new Luttinger liquid in terms of the fields $\phi_M(\vec{r})$, $\theta_M(\vec{r})$ with parameter $K_M = 4K$.

To calculate the dc spin current through the system we use~\cite{Mei03} that in linear-response theory the dc spin current is given by $I_s = G \Delta B$, where the conductance $G$ is given by
\begin{equation}
G = -i\frac{(g \mu_B)^2}{ \pi^2 \hbar }\lim_{\omega \to 0}\left[\omega\left.G_{\phi\phi}(x,x',\omega_n)\right|_{i\omega_n \to \omega + i \eps}\right],
\end{equation}
and $G_{\phi\phi}(x,x',\omega_n)$ is the time-ordered Green's function in imaginary time
\begin{equation}
G_{\phi\phi}(x,x',\omega_n) = \int_0^{\beta} \ud \tau e^{i\omega_n\tau}\left\la T_{\tau}\phi(x,\tau)\phi(x',0)\right\ra_{S_0}.
\end{equation}
Here $\omega_n$ is the Matsubara frequency. At zero temperature, and assuming infinitesimal dissipation in the reservoirs, the $\omega \to 0$ limit of this Green's function can be determined for the inhomogeneous system including the two reservoirs,~\cite{Mas95} and is given by $\frac{K_r}{2|\omega_n|}$. The effect of the entire mapping of the original sine-Gordon model onto the new free Luttinger liquid can be captured here by replacing $g\mu_B \to g\mu_B/2$ and $K_r \to 4K_r$, so that the conductance in the conducting phase is $G = K_r\frac{(g \mu_B)^2}{h}$. Following Ref. \onlinecite{Sta98} we conclude that excitations that are injected at energies above the Mott gap are transported through the chain, giving rise to the aforementioned conductance, whereas excitations injected at chemical potential lower than the gap are fully reflected. Assuming that $g \mu_B B_0 \approx M$, this gives rise to the spin current
\begin{equation}
I_s(\Delta B) = K_r\frac{(g \mu_B)^2}{h} \Delta B \Theta(-\Delta B),
\end{equation}
where $\Theta(-\Delta B)$ is the unit step function. Since spin transport is absent for $\Delta B > 0$, we have $I_r (\Delta B)= I_s(\Delta B)/2$. In these calculations we assumed that the length of the spin chain $L \to \infty$, so that we can neglect tunneling of solitons, and we have assumed zero temperature. Finite size and temperatures are known to give corrections to the conductance.~\cite{Pon98,Gar08}

\section{Suppressed anisotropy}
\label{sec:WC}
\label{sec:wc}
To determine the rectification current resulting from the exchange anisotropy as given in Eq. (\ref{eq:scen2}) we need to calculate the spin current in the system given the action $S[\phi] = S_0[\phi] + S_\textrm{I}[\phi]+S_B[\phi]$. For simplicity we assume here that the impurity is located at $x_0=0$. We ignore the bulk umklapp scattering term, since we have shown in Sec. $\ref{sec:Ren}$ that the contribution of this term is irrelevant for the parameters used here. From the RG analysis it also followed that the most important terms are the offset-induced backscattering terms and that in regions where $B_0 \approx \Delta B/2$ the effect of the terms describing combined forward- and backscattering and the offset-induced backscattering terms can become comparable, due to the extra $\sigma$ in front of the latter terms. Therefore we need to calculate the spin current due to both types of contribution. We will show that the rectification effect appears in the contributions to the spin current that are second order in the coupling constants.

We calculate the spin current using the Keldysh technique.~\cite{Kel65} We assume that at $t=-\infty$ the system is described by the action $S_0[\phi]+S_B[\phi]$, and that the perturbation $S_{\textrm{I}}[\phi]$ is turned on adiabatically. From conservation of spin it follows that the expression for the spin current in the Luttinger liquid is given by $I_s(\vec{r}) = -\frac{g \mu_B}{\pi} \partial_t \phi(\vec{r})$. The spin current can then be calculated as~\cite{Dolc05}
\begin{eqnarray}
I_s = -\frac{g \mu_B}{\pi} \partial_t \frac{1}{2}\sum_{\eta = \pm} \left\la \phi^{\eta}(\vec{r})\right\ra_S = \frac{g \mu_B }{\pi} i\partial_t \left(\frac{\delta Z[J(\vec{r})]}{\delta J(\vec{r})}\right).
\label{eq:curZ}
\end{eqnarray}
Here $\left\la \phi^{\pm}(\vec{r})\right\ra_S$ is the average of the field $\phi(\vec{r})$ over the Keldysh contour with respect to the action $S[\phi]$, where the $\pm$ denotes that the field is located on the positive, respectively negative, branch of the contour. The right-hand side contains the functional derivative of the partition function of the system with respect to the generating functional $J(\vec{r})$, which is given in Eq. (\ref{eq:partfun}). The details of the calculation are given in Appendix \ref{ch:A1}; here we summarize the results. We find that there are two fundamentally different contributions to the spin current,
\begin{equation}
I_s(\Delta B) = I_0(\Delta B) + I_{\textrm{BS}}(\Delta B).
\end{equation}
Here, $I_0(\Delta B)$ is the spin current through the systems in the absence of $S_I[\phi]$, given by the well-known expression 
\begin{equation}
I_0(\Delta B) = K_r\frac{(g \mu_B)^2}{h} \Delta B,
\end{equation}
and $I_{\textrm{BS}}(\Delta B)$ describes (negative) contributions to the spin current due to $S_{\textrm{I}}[\phi]$, which we will derive next.

The contribution to the spin current resulting from a $\sigma\frac{\lambda_3^a}{\pi^2}\cos[2\phi(0,t)]$ term (which describes offset-induced backscattering) is given by
\begin{equation}
I_{3a} (\Delta B) = \frac{ g \mu_B \sigma^2\left(\lambda_{3,R}^a\right)^2 }{ \pi^4 }\frac{1}{a_s} K_r A_0(\Delta B),
\label{eq:IBS3aM}
\end{equation}
where
\begin{equation}
A_0(\Delta B) = -\frac{K4^K\pi}{\Gamma(1+2K)}\gamma_R \left|\gamma_R\right|^{-2+2K}e^{-2|\gamma_R|}.
\label{eq:A0M}
\end{equation}
Here we introduce the dimensionless parameters $\lambda_{3,R}^a = \lambda_3^a/(\hbar \omega_c)$ and $\gamma_R = K_r g\mu_B \Delta B/(\hbar \omega_c)$. As before, $\omega_c$ denotes the UV cutoff of our theory, given by $\omega_c = u / a $. In Sec. \ref{sec:Ren} it was determined that the backscattering term scales as $1-K$ under renormalization. To improve our result we should therefore not use the bare coupling $\lambda_{3,R}^a$, but instead the renormalized coupling. Since we assumed an infinitely long chain, and consider zero temperature, the renormalization group procedure is stopped on the energy scale determined by the magnetic field, $g \mu_B \Delta B$. We can account for this by replacing $\lambda'^a_{3,R} \to \left|\frac{g \mu_B \Delta B}{\hbar \omega_c}\right|^{-1+K} \lambda^a_{3,R}$. At this point we have determined the spin current resulting from the backscattering term. By repeating the previous calculation with both the $\lambda_{3}^a$- and the $\lambda_{3}^b$-proportional terms included it is easily seen that the $\left(\lambda_{3}^b\right)^2$-proportional contribution to the spin current is also given by Eq. (\ref{eq:IBS3aM}), with $\left(\lambda_{3,R}^a\right)^2$ replaced by $\left(\lambda_{3,R}^b\right)^2$. The two cross terms proportional to $\lambda_{3}^a\lambda_{3}^b$ do not contribute to the spin current, since they cancel each other.

The calculation of the contribution to the spin current due to a term $\frac{a\lambda_4^a }{\pi^2}\partial_x\phi(0,t) \cos 2\phi(0,t)$ that describes combined forward- and backscattering proceeds along the same lines; we refer again to Appendix \ref{ch:A1} for the details. The resulting contribution to the spin current is given by
\begin{equation}
I_{4a}(\Delta B) = \frac{g\mu_B\left(\lambda_{4,R}^a\right)^2}{\pi^4}\frac{1}{a_s}K_rA_1(\Delta B),
\label{eq:IBS4aM}
\end{equation}
where
\begin{equation}
A_1(\Delta B) = -\frac{4^{K}\pi}{\Gamma(2+2K)}\gamma_R \left|\gamma_R\right|^{2K}e^{-2|\gamma_R|}.
\label{eq:A1M}
\end{equation}
Again, according to the RG analysis, we have to replace the bare coupling with its renormalized value, in this case, $\lambda'^a_{4,R} \to \left|\frac{g \mu_B \Delta B}{\hbar \omega_c}\right|^{-1+K} \lambda_{4,R}^a$. Like with the backscattering terms, we can easily determine the effect of the $\lambda_{4}^a$ and $\lambda_{4}^b$ terms combined. The spin current proportional to  $\left(\lambda_{4}^b\right)^2$ is given by Eq. (\ref{eq:IBS4aM}) with $\left(\lambda_{4,R}^a\right)^2$ replaced by $\left(\lambda_{4,R}^b\right)^2$, and the two cross terms cancel each other.

Finally, we need to consider the cross terms between the $\lambda_3$  and $\lambda_4$ terms, such as, for instance, $\lambda_{3}^a\lambda_{4}^b$. Using the results from Appendix \ref{app:corr} it is easily seen that these always vanish, since they are all proportional to $\left\la T_c\partial_x \phi^{\eta}(0,t)e^{\pm 2i \left[\phi^{\eta}(0,t)-\phi^{\eta'}(0,t')\right]}\right\ra = 0$. The total backscattered current is then given by
\begin{equation}
I_{\textrm{BS}} (\Delta B)= \frac{g \mu_B}{\pi^4}\frac{K_r}{a_s} \left\{\sigma^2\lambda_{3,\textrm{eff}}^2 A_0(\Delta B) +\lambda_{4,\textrm{eff}}^2 A_1(\Delta B)\right\},
\label{eq:IBStot}
\end{equation}
where $\lambda_{i,\textrm{eff}} = \left[\left(\lambda'^a_{i,R}\right)^2+ \left(\lambda'^b_{i,R}\right)^2\right]^{1/2}$. 

Equation (\ref{eq:IBStot}) is the main result of this section. From the explicit form of $A_0(\Delta B)$ and $A_1(\Delta B)$ [see Eqs. (\ref{eq:A0M}) and (\ref{eq:A1M})] it is clear that the impurity flows to strong coupling for low $\Delta B$, as was expected from the RG analysis. We note that, since both $A_0(\Delta B)$ and $A_1(\Delta B)$ are odd in $\Delta B$, one could naively think that the impurity does not contribute to the rectification current, which requires the spin current to have a component that is even in $\Delta B$. However, from the explicit form of the $\lambda_{i,\textrm{eff}}$ [see Eq. (\ref{eq:prefac}) for the bare couplings] it follows that these couplings also contain parts that are proportional to $\Delta B$. Furthermore, the part of the spin current that is proportional to $A_0(\Delta B)$ is proportional to $\sigma^2$, which also contains a $\Delta B$. Physically, these terms are caused by the fact that an incoming excitation sees a slightly different impurity depending on the energy with which it comes in. The spin current Eq. (\ref{eq:IBStot}) therefore has components even in $\Delta B$, which contribute to the rectification current. We are now also in a position to explain why it is needed to move the system away from half filling in order to obtain a nonzero rectification current. If we set $B_0 = 0$ in Eq. (\ref{eq:prefac}), it turns out that the $\lambda_{i,\textrm{eff}}$'s and $\sigma$ contain only terms that are even in $\Delta B$, so that the spin current is odd in $\Delta B$. In contrast, when $ B_0 \neq 0$, we create terms in the $\lambda_{i,\textrm{eff}}$'s and $\sigma$ that are odd in $\Delta B$, thereby causing a nonzero rectification current.

\section{Estimate of rectification currents}
\label{sec:Est}
\begin{figure}
\centering
\includegraphics[width=1.0\columnwidth]{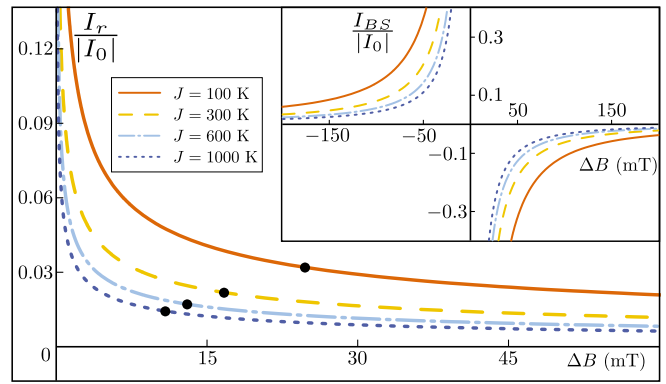}
\caption[]{(Color Online) Normalized rectification current $I_\textrm{r}/|I_0|$ (main figure) and backscattering current $I_\textrm{BS}/|I_0|$ (inset) as a function of the applied magnetic-field difference $\Delta B$ for different values of the exchange interaction $J$. Parameters are $K=0.63$, $B_0 = 750$ mT, and $\Delta' = 0.5$. The black dots in the main figure denote the values $\Delta B^*$ for which $\textrm{max}\left[|I_{\textrm{BS}}(\Delta B^*)|,|I_{\textrm{BS}}(-\Delta B^*)|\right] = |I_0(\Delta B^*)|$. As explained in the text, our perturbative results are not valid for $|\Delta B| < |\Delta B^*|$.}
\label{fig:Jdep}
\end{figure}
In this section we will give numerical estimates of the rectification current for realistic experimental parameters. Experimental results~\cite{Ten952} show that the exchange coupling in certain spin chains can be on the order of $J \approx 10^2$ K. The effect of a different exchange interaction is illustrated in Fig. \ref{fig:Jdep}. In the analysis of these results one must keep in mind that, since our perturbative calculation of $I_{\textrm{BS}}(\Delta B)$ diverges for $\Delta B \to 0$, it breaks down for field differences $|\Delta B| < |\Delta B^*|$, where $\Delta B^*$ is the field such that $\textrm{max} \left[I_{\textrm{BS}}(\Delta B^*),I_{\textrm{BS}}(-\Delta B^*)\right] = I_0(\Delta B^*)$. Instead of showing the apparent divergent behavior. the backscattering current must go to zero for $\Delta B < \Delta B^*$. With this in mind, Fig. \ref{fig:Jdep} shows that a smaller exchange interaction $J$ gives rise to a larger rectification current at equal $\Delta B$. The maximum value of $I_r(\Delta B)$ will also be reached at a higher value of $\Delta B$. In order to get the largest possible rectification current it is thus required to use a material with an exchange interaction as small as possible, with the constriction that it must be large enough to yield its maximum at reasonable values of $\Delta B$.

\begin{figure}
\centering
\includegraphics[width=1.0\columnwidth]{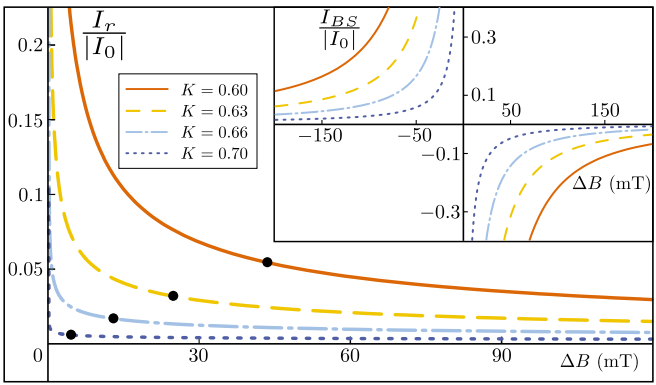}
\caption[]{(Color Online) $I_\textrm{BS}/|I_0|$ as a function of the applied magnetic field difference $\Delta B$ for different values of $K$. Parameters are $B_0=750$ mT, $J = 100$ K, and $\Delta' = 0.5$. See the caption of Fig. \ref{fig:Jdep} for the explanation of the black dots.}
\label{fig:Kdep}
\end{figure}

In Fig. \ref{fig:Kdep} we show the dependence of the backscattering and rectification current on the Luttinger liquid parameter $K$. The behavior for smaller $K$ is similar to the behavior shown for smaller $J$: The maximum rectification current is increased, but occurs at a higher value of $\Delta B$. Indeed, we know that both $I_s(\Delta B)$ and $I_r(\Delta B)$ obey a power-law dependence of $\Delta B$ with negative exponent. Since the modulus of this exponent decreases for increasing $K$, the behavior is as expected.
\begin{figure}
\centering
\includegraphics[width=1.0\columnwidth]{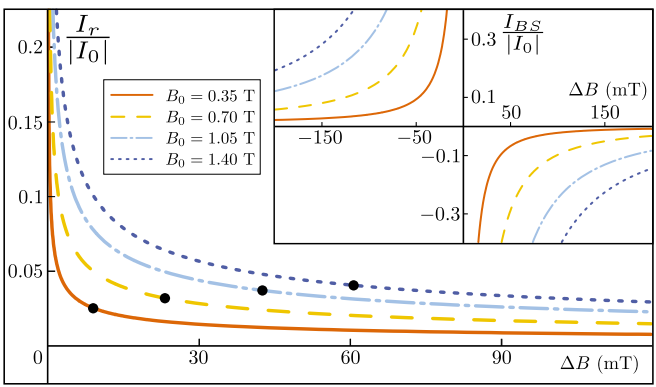}
\caption[]{(Color online) $I_\textrm{BS}/|I_0|$ as a function of the applied magnetic-field difference $\Delta B$ for different values of $B_0$. Parameters are $K=0.63$, $J = 100$ K, and $\Delta' = 0.5$. See the caption of Fig. \ref{fig:Jdep} for the explanation of the black dots.}
\label{fig:gammadep}
\end{figure}

Figure \ref{fig:gammadep} shows the dependence of the backscattering current on the applied magnetic field $B_0$. A larger amount of doping clearly corresponds to larger possible rectification current, again at the price of a higher required magnetic field $\Delta B$. This can be understood by realizing that $B_0$ determines to a large extent how much of the impurity the incoming excitations see at low energies $\Delta B \ll B_0$, as follows from the $\sigma$ dependence in the $\lambda_3^{a,b}$ terms in the action Eq. (\ref{eq:SI}).

In Ref. \onlinecite{Mei03} it has been shown that $N$ parallel uncoupled AF spin chains connected between two three-dimensional AF spin reservoirs, each one carrying a spin current $I_s(\Delta B)$, give rise to an electric field
\begin{equation}
\vec{E}(\vec{r}) = N \frac{\mu_0}{2\pi}\frac{I_S(\Delta B)}{r^2}\left(0, \cos 2\phi, -\sin 2\phi\right)
\label{eq:Efield}
\end{equation}
Here the spin chains are assumed to extend in the $x$ direction, and $z$ is the quantization axis as before. Also, $r = \sqrt{y^2+z^2}$, $\sin \phi = y/r$, and $\cos \phi = z/r$. We can use this electric field to measure the rectification current $I_r(\Delta B)$. To illustrate the method we assume $K = 0.6$, $B_0 = 750$ mT, and $N = 10^4$. We apply the time-dependent field $\Delta B (t) = \Delta B \cos \left( \omega t\right)$ to the left reservoir. From Eq. (\ref{eq:IBStot}) it follows that, if we trust our perturbative calculation of $I_{\textrm{BS}}$ up to the value $\Delta B^* \approx -43$ mT, where $|I_0(\Delta B^*)| \approx |I_{\textrm{BS}}(\Delta B^*)|$, the difference in magnitude between $|I_S(\Delta B^*)|$ and $|I_S(-\Delta B^*)|$ is on the order of 10\% of the unperturbed current $|I_0(\Delta B^*)|$. From Eq. (\ref{eq:Efield}), and assuming $r=10^{-5}$,  it then follows that the difference in voltage drop between two points $\left(0,r,0\right)$ and $\left(0,0,r\right)$ is $V_S\approx 10^{-13}$ V, which is within experimental reach. Here we note that as long as the driving frequency $\omega \ll J/\hbar$, our calculation of the spin current in the dc limit remains valid. Another constraint is given by the spin-relaxation time $\tau_s$, which is typically on the order of $10^{-7}$ s.~\cite{Zha98} This allows us to use frequencies in the MHz-GHz range.

Another possibility to observe the rectification effect is by spin accumulation. By applying again an ac driving field to the left reservoir it is possible to measure an accumulation of spin in the right reservoir, since transport is asymmetric with respect to the sign of $\Delta B$. We consider again $10^{4}$ parallel spin chains with $K = 0.6$, consider $B_0 = 750$ mT, and an amplitude $\Delta B = 43$ mT for the driving field. For $\Delta B (t)< 0$ the spin current is always zero. For $\Delta B (t)> 0$ there is a nonzero spin current. Assuming that the spin current is $10 \%$ of the unperturbed spin current [the value at $\Delta B (t) = \Delta B$] over the entire range $\left( 0, \Delta B \right)$, the rate of spin accumulation is about $ 10^{-11}$ $\textrm{J} \textrm{T}^{-1} \textrm{s}^{-1}\approx 10^{12}$ magnons per second. We also note that, contrary to the electric case, where charge repulsion prevents a large charge accumulation, there is no strong mechanism that prevents spin accumulation in this scenario.

\section{Conclusions}
\label{conclusion}
In this work we have proposed various realizations of the rectification effect in spin chains, consisting of either ferromagnetically or antiferromagnetically coupled spins, adiabatically connected to two spin reservoirs. For both ferromagnetic and  antiferromagnetic spin chains we find that the two crucial ingredients to achieve a nonzero rectification current are an anisotropy in the exchange interaction in combination with an offset magnetic field, both in the $z$ direction. An enhanced anisotropy creates a gap in the excitation spectrum of the spin chain. Using the  magnetic field to tune the chemical potential of the excitations to just below the lower edge of the gap then allows us to achieve a large rectification current for realistic values of the applied magnetic field. For antiferromagnetic coupling and suppressed exchange interaction in the $z$ direction we find that a uniform anisotropy is not sufficient to achieve a sizable rectification effect; instead we use a spatially varying anisotropy, which we attain in the form of a site impurity. Away from half filling this impurity flows to strong coupling under renormalization, which allows us to achieve a sizable rectification effect for realistic values of the applied magnetic field. We have also proposed several ways to observe this rectification effect.
\section{Acknowledgments}
We thank B. Braunecker for valuable discussions. This work has been supported by the Swiss NSF and NCCR Nanoscience Basel.
\appendix
\section{Determining Luttinger Liquid parameters from the nonlinear sigma model.}
\label{ch:A0}
In this section we calculate the magnetic susceptibility of a three-dimensional antiferromagnet, denoted by $\chi^{\textrm{3D}}_{\alpha \beta}(q)$. In the continuum limit, the magnetic susceptibility is defined as
\begin{equation}
 \chi^{\textrm{3D}}_{\alpha\beta}(q) = \left(\frac{g \mu_B}{a^3}\right)^2 \frac{1}{\hbar}\int \ud^{4} s  e^{i q \cdot s} \left\la T_{\tau} S_\alpha(s)S_\beta(0)\right\ra,
\label{eq:chi}
\end{equation}
where $s = (\tau,\vec{r})^T$, $q = (\omega_n, \vec{k})^T$, and $q\cdot s = \omega_n\tau - \vec{k}\cdot \vec{r}$. 

We start from the isotropic Heisenberg model in three dimensions, described by Eq. (\ref{eq:Hamil}) with $\Delta_{ij} = 1$ and $B_i = 0$. We introduce the unit vector field $\vec{n}^s_i = \vec{S}_i/S$. This vector can be written as a staggered N\'eel vector part $\vec{n}_i$ plus a fluctuating part $\vec{l}_i$, both of which are slowly varying on the scale $a$,
\begin{equation} 
 \vec{n}^s_i = (-1)^ i\vec{n}_i\sqrt{1-a^2\vec{l}^2} + a\vec{l}_i \approx (-1)^i \vec{n}_i + a \vec{l}_i.
\label{eq:Neel}
\end{equation}
In the continuum limit it can be shown~\cite{All97} that the imaginary time action describing this system can be approximated as $S[\vec{n},\vec{l}] = S_0[\vec{n},\vec{l}] + S_l[\vec{n},\vec{l}]$, where 
\begin{eqnarray}
 S_0[\vec{n},\vec{l}] & = & \frac{\hbar}{2\gamma} \int \ud^4 s \left[\frac{1}{c}\left(\partial_\tau \vec{n}\right)^2 + \sum_{i = 1}^3 c \left(\partial_{r_i} \vec{n}\right)^2\right],
\label{eq:appS0} \\
S_l[\vec{n},\vec{l}] & = & 12\hbar \alpha \int \ud^4 s \left[\vec{l} + i\frac{1}{48a^2\alpha}\left(\vec{n}\times \partial_\tau \vec{n}\right)\right]^2.
\label{eq:appSl}
\end{eqnarray}
Here, $\alpha = J/(8a\hbar)$, $c = \sqrt{3}Ja/\hbar$, and $\gamma = 4\sqrt{3}a^2$. From a substitution of Eq. (\ref{eq:Neel}) in Eq. (\ref{eq:chi}) it is clear that we can write
\begin{equation}
 \chi^{\textrm{3D}}_{\alpha\beta}(q) = \left(\frac{g \mu_B}{2 a^2}\right)^2 \frac{1}{\hbar}\int \ud^{4} s  e^{i q \cdot s} \left\la T_{\tau} l_\alpha(s)l_\beta(0)\right\ra_S,
\label{eq:xhi2}
\end{equation}
where the Green's function is calculated with respect to the action $S[\vec{n},\vec{l}]$. The Green's function is determined from the partition function $Z = \int D \vec{n}D\vec{l}\delta(\vec{n}^2-1)\exp\left(-\hbar^{-1}S[\vec{n},\vec{l}]\right)$. We can perform the functional integration over $ \vec{l}$ in Eq. (\ref{eq:xhi2}) to arrive at
\begin{eqnarray}
 \chi^{\textrm{3D}}_{\alpha\beta}(q) & = & \left(\frac{g\mu_B }{\gamma }\right)^2 \frac{1}{\hbar c} \bigg[ \gamma \delta_{\alpha\beta} -\int \ud^4 s e^{-q\cdot s} \nonumber\\
& & \times \left\la \left(\vec{n}\times\partial_\tau\vec{n}\right)_\alpha(s)\left(\vec{n}\times\partial_\tau\vec{n}\right)_\beta(0) \right\ra_{S_0}\bigg],
\label{eq:chi3}
\end{eqnarray}
where we replaced $\tau \to c \tau$. Here, the average is with respect to $S_0[\vec{n},\vec{l}]$. Assuming the magnetization is ordered along the $z$ axis, we write $\vec{n} = \left(\sigma_1,\sigma_2,\Pi\right)^T$. The $\sigma_i$'s describe thermal fluctuations perpendicular to the direction of order,~\cite{Cha95} and $\Pi = \sqrt{1-\vec{\sigma}^2}$. Also, $\vec{\sigma} = (\sigma_1,\sigma_2)^T$. For low temperatures, we can then rewrite the action of Eq. (\ref{eq:appS0}) as
\begin{equation}
S_0[\vec{\sigma}] =   -\frac{\hbar}{2\gamma} \int \ud^4 s \vec{\sigma}\cdot \left[\partial^2_\tau+\sum_{i=1}^3 \partial_{x_i}^2\right]\vec{\sigma}.
\end{equation}
From this and Eq. (\ref{eq:chi3}) it then follows that the magnetic susceptibility in the gapless direction is given by
\begin{equation}
 \chi^{\textrm{3D}}(q) = \frac{\left(g \mu_B\right)^2}{\hbar} \frac{c}{\gamma} \frac{\vec{q}^2}{\omega_n^2 + (c\vec{q})^2}.
\end{equation}
To compare this relation to the magnetic susceptibility in one dimension, we use $\chi^{\textrm{1D}}(q) = a^2 \chi^{\textrm{3D}}(q)$. Since we know~\cite{Gia03} that in the Luttinger liquid model
\begin{equation}
 \chi^{\textrm{LL}}_{zz}(q) = \frac{(g\mu_B)^2}{\hbar}\frac{u K }{\pi} \frac{\vec{q}^2}{(u\vec{q})^2 + \omega_n^2},
\end{equation}
we have
\begin{equation}
u_r = \sqrt{3}Ja/\hbar \qquad K_r = \pi/(4\sqrt{3})
\end{equation}
as effective Luttinger liquid parameters of the three-dimensional reservoirs.
\section{Calculation of the spin current in the Keldysh formalism}
\label{ch:A1}
In this appendix we will show how to calculate the spin current $I_s(\Delta B)$ in a system described by the action $S[\phi] = S_0[\phi] + S_\textrm{I}[\phi]+S_B[\phi]$. According to Eq. (\ref{eq:curZ}), the spin current is given by
\begin{eqnarray}
I_s = -\frac{g \mu_B}{\pi} \partial_t \frac{1}{2}\sum_{\eta = \pm} \left\la \phi^{\eta}(\vec{r})\right\ra_S = \frac{g \mu_B i }{\pi} \partial_t \left(\frac{\delta Z[J(\vec{r})]}{\delta J(\vec{r})}\right).
\end{eqnarray}
We start by deriving the contribution to the current due to a $\sigma \frac{\lambda_3^a}{\pi^2} \cos 2\phi(0,t)$ term. In this case the partition function can be written in vectorized form as
\begin{eqnarray}
Z[J(\vec{r})] &=& \int D \vec{\phi} \exp \bigg[ -\frac{1}{2} \int \ud^2 r \ud^2 r' \nonumber\\
 &&\times\left[ \vec{\phi}(\vec{r})\cdot G_0(\vec{r},\vec{r}')\vec{\phi}(\vec{r}') - 2i\delta(\vec{r}-\vec{r}')\vec{\phi}(\vec{r})\cdot Q \vec{J}(\vec{r}')\right]\bigg] \nonumber\\
 &&\times\exp \left[\frac{i\sigma\lambda_3^a}{\pi^2\hbar}\int_{-\infty}^{\infty} \ud t \sum_{\eta = \pm} \eta \cos \left[2 \phi^{\eta}(0,t)\right]\right],
\label{eq:partfun}
\end{eqnarray}
where we have defined
\begin{equation}
\begin{split}
\vec{\phi}(\vec{r}) &=  \left(\begin{array}{c} \phi^+(\vec{r}) \\ \phi^-(\vec{r}) \end{array}\right), \\
\vec{J}(\vec{r}) &= \sqrt{2}\left(\begin{array}{c} -\frac{g \mu_B}{\pi \hbar}\partial_x B(\vec{r}) \\ J(\vec{r})/2 \end{array}\right), \\
Q &= \frac{1}{\sqrt{2}} \left(\begin{array}{cc} 1 & -1 \\ 1 & 1 \end{array}\right), \\
G_0(\vec{r},\vec{r}') & =  \left(\begin{array}{cc} G_0^{++}(\vec{r},\vec{r}') & G_0^{+-}(\vec{r},\vec{r}') \\ G_0^{-+}(\vec{r},\vec{r}') & G_0^{--}(\vec{r},\vec{r}') \end{array}\right). \end{split}
\end{equation}
Here $G_0^{\eta,\eta'}(\vec{r},\vec{r}')$ is the contour-ordered Green's function
\begin{equation}
G_0^{\eta,\eta'}(\vec{r},\vec{r}') = \left\la T_c \phi^{\eta}(\vec{r})\phi^{\eta'}(\vec{r}')\right\ra_{S_0}.
\end{equation}
We will also need the retarded, advanced, and Keldysh Green's functions $G_0^{R/A/K}(\vec{r},\vec{r}')$, which are related to the contour-ordered Green's functions in the usual way. Performing the linear transformation $\vec{\phi}'(\vec{r}) = \vec{\phi}(\vec{r}) - i \int \ud \vec{r}'G_0(\vec{r},\vec{r}') Q^T \vec{J}(\vec{r}')$ shows that the partition function can be factorized into two parts, the first one containing the quadratic part of the action, the second containing the backscattering term. The spin current is therefore made up of two contributions. The first, $I_0$,  comes from the quadratic action, and describes the spin current in the absence of any backscattering. The second, $I_{3a}$, originates from the backscattering term. Using the explicit form for the magnetic field Eq. (\ref{eq:mag}) it turns out that the expression for the unperturbed spin current is given by
\begin{equation}
I_0 = i \frac{(g \mu_B)^2}{\pi^2 \hbar} \Delta B \partial_t \int_{-\infty}^{\infty} \ud t'G_0^R(\vec{r};-L/2,t').
\end{equation}
Since this is exactly the expression for the spin current in the linear-response regime,~\cite{Mei03} which yields $I_0 = K_r\frac{(g \mu_B)^2}{h}\Delta B$, we can infer that
\begin{equation}
\int_{-\infty}^{\infty} \ud t'i G_0^R(\vec{r};-L/2,t') = \frac{\pi K_r t}{2}.
\label{eq:eqcur}
\end{equation}
To calculate the contribution from the backscattering term we perform the functional derivative of the partition function with respect to $J(\vec{r})$. The resulting expression is
\begin{eqnarray}
I_{3a} & = & \frac{ 2 g \mu_B \sigma \lambda_3^a }{ \pi^3 \hbar } \partial_t \int_{-\infty}^{\infty} \ud t' i G_0^R(\vec{r};0,t') \nonumber\\
& & \times\left\la \sin\left[2\phi'^+ (0,t') - 2\gamma t'\right]\right\ra_{S},
\end{eqnarray}
where $\gamma = K_r \frac{g \mu_B \Delta B}{\hbar}$. Expanding this term in $\lambda_3^a$ gives
\begin{eqnarray}
I_{3a} &=& i \frac{ 2 g \mu_B \sigma^2\left(\lambda_3^a\right)^2 }{ \pi^5 \hbar^2 } \partial_t \int_{-\infty}^{\infty} \ud t' i G_0^R(\vec{r};0,t') \ud t''  \times\nonumber\\
&&\sum_{\eta = \pm} \eta \left\la \sin \left[2\phi'^+(0,t')-2\gamma t'\right] \cos\left[2\phi'^{\eta}(0,t'')-2\gamma t''\right]\right\ra_{S_0}.\nonumber
\end{eqnarray}
Since contour ordering reduces to time ordering for two fields on the same branch of the Keldysh contour, and because the system in equilibrium is time-reversal invariant, the term with the fields at times $t'$ and $t''$ on the same branch averages to zero. Switching the variable of integration to $\tau = t'-t''$ and using Eq. (\ref{eq:eqcur}) it is seen that the contribution from the backscattering term is given by
\begin{eqnarray}
I_{3a} &=& \frac{ g \mu_B \sigma^2\left(\lambda_{3,R}^a\right)^2 }{ \pi^4 }\frac{1}{a_s} K_r A_0(\Delta B).
\label{eq:IBS3a}
\end{eqnarray}
Here $a_s = a/u$ can be interpreted as the cutoff in the time domain. As before, we identify the cutoff in time with the cutoff in frequency: $a_s = \omega_c^{-1}$, so that we can write the expression for the backscattering current in terms of the dimensionless parameters $\lambda_{3,R}^a = \lambda_3^a/(\hbar \omega_c)$, $\gamma_R = K_r g\mu_B \Delta B/(\hbar \omega_c)$, and $\bar{\tau} = \tau/a_s$. Here
\begin{equation}
A_0(\Delta B) = \frac{i}{2}\int_{-\infty}^{\infty} \ud \bar{\tau}\sin \left[2 \gamma_R\bar{\tau}\right] e^{-2\left\la T_c\left[\phi^+(0,\tau)-\phi^-(0,0)\right]^2\right\ra_{S_0}}.
\label{eq:A0}
\end{equation}
The contour-ordered correlation function can be determined from the inhomogeneous Luttinger liquid model~\cite{Mas95} and the fluctuation-dissipation theorem, and can be written in terms of $\bar{\tau}$ as
\begin{equation}
\left\la T_c\left[\phi^+(0,\tau) - \phi^-(0,0)\right]^2\right \ra_{S_0} = K\ln \left[-i\bar{\tau} + 1\right].
\end{equation}
With this expression we can rewrite Eq. (\ref{eq:A0}) as a function of $\gamma_R$ and $K$ only
\begin{equation}
A_0(\Delta B) = -\frac{K4^K\pi}{\Gamma(1+2K)}\gamma_R \left|\gamma_R\right|^{-2+2K}e^{-2|\gamma_R|}.
\end{equation}
Next we discuss the contribution of the $\frac{a\lambda_4^a }{\pi^2}\partial_x\phi(0,t) \cos 2\phi(0,t)$ term from Eq. (\ref{eq:SI}). We again perform the linear transformation $\vec{\phi}(\vec{r}) \to \vec{\phi}'(\vec{r})$, which yields
\begin{eqnarray}
I_{4a} &=& \frac{ 2g \mu_B \lambda_4^a a }{ \pi^3 \hbar } \partial_t \int_{-\infty}^{\infty} \ud t' i G_0^R(\vec{r};0,t') \nonumber\\
&& \times\left\la \partial_{x'} \phi'^+(0,t') \sin\left[2\phi'^+ (0,t') - 2\gamma t'\right]\right\ra_{S}.
\label{eq:IBS}
\end{eqnarray}
There is also an additional term in the expression for the spin current introduced by the linear transformation, which is proportional to $\partial_x \int \ud t G_0^R(0,t;\vec{r}'')$. Equation (\ref{eq:eqcur}), however, shows that this term is zero. We can expand Eq. (\ref{eq:IBS}) in $\lambda_4^a$,
\begin{eqnarray}
I_{4a} &=& i \frac{ 2 g \mu_B \left(\lambda_4^a\right)^2 a^2}{ \pi^5 \hbar^2 } \partial_t \int_{-\infty}^{\infty} \ud t' i G_0^R(\vec{r};0,t') \ud t'' \nonumber\\
&&\times  \sum_{\eta = \pm} \eta \big\la \partial_{x'} \phi'^+(0,t') \sin \left[2\phi'^+(0,t')-2\gamma t'\right] \nonumber\\
& & \times \partial_{x''}\phi'^{\eta}(0,t'') \cos\left[2\phi'^{\eta}(0,t'')-2\gamma t''\right]\big\ra_{S_0}.
\end{eqnarray}
We now use the results from Appendix \ref{app:corr} to evaluate the correlation function,
\begin{equation}
\partial_{x}\partial_{x'} \left\la T_c \phi^+(0,t)\phi^-(0,t')\right\ra_{S_0} = \frac{2}{(ua_s)^2}\left(\frac{1+i\bar{\tau}}{1+\bar{\tau}^2}\right)^2.
\end{equation}
We also use that $\partial_{x} \left\la T_c \phi^+(0,t)\phi^-(0,t')\right\ra_{S_0} = 0$, so that the spin current is given by
\begin{equation}
I_{\textrm{BS}} = \frac{g\mu_B\left(\lambda_{4,R}^a\right)^2}{\pi^4}\frac{1}{a_s}K_rA_1(\Delta B),
\label{eq:IBS4a}
\end{equation}
where the dependence on the parameters of the spin chain and the applied magnetic field difference is contained in the function $A_1(\Delta B)$
\begin{eqnarray}
A_1(\Delta B) &=& i\frac{(u a_s)^2}{2} \int_{-\infty}^{\infty} \ud \bar{\tau} e^{-2\left\la T_c\left[\phi^+(0,\tau)-\phi^-(0,0)\right]^2\right\ra_{S_0}}\nonumber\\
&& \times \left\la \partial_x \phi'(0,\tau)\partial_x\phi'(0,0)\right\ra_{S_0}\sin \left[2\gamma_R \bar{\tau}\right],
\end{eqnarray}
where we again used the fact that terms containing correlators with both fields on the same branch of the contour vanish. Straightforward manipulations show that $A_1(\Delta B)$ is given by
\begin{equation}
A_1(\Delta B) = -\frac{4^{K}\pi}{\Gamma(2+2K)}\gamma_R \left|\gamma_R\right|^{2K}e^{-2|\gamma_R|}.
\label{eq:A1}
\end{equation}
\section{Correlators}
\label{app:corr}
In this appendix we show how to calculate a correlation function of the form
\begin{equation}
\left\la T_c\Pi_i\partial_{x_i}\phi(x_i,t_i)e^{i\sum_jA_j\phi(x_j,t_j)}\right\ra_{S_0}.
\label{eq:corr2}
\end{equation}
First of all, we can pull the derivatives outside of the expectation value, so we can calculate $\left\la T_c \Pi_i\phi(x_i,t_i)e^{i\sum_jA_j\phi(x_j,t_j)}\right\ra_{S_0}$ instead and take the derivatives afterward. For a lighter notation we prove the required results in discrete space, making the connection $\phi_i = \phi(\vec{r}_i)$. We use the main result of Gaussian integration~\cite{Atla06}
\begin{equation}
\int \ud \vec{\phi} e^{-(1/2) \vec{\phi}^T\cdot\bar{G}\vec{\phi} + \vec{j}^T\cdot \vec{\phi}} \propto  e^{\vec{j}^T\cdot\bar{G}^{-1}\vec{j}}.
\label{eq:gauss}
\end{equation}
Given the partition function
\begin{equation}
Z = \int \ud \vec{\phi} e^{-(1/2)\vec{\phi}^T\cdot\bar{G} \vec{\phi}},
\end{equation}
we can then calculate, for instance, $\left\la \phi_i\phi_j\right\ra_{S_0}$ by letting $\left.\partial^2_{j_n j_m}\right|_{\vec{j}=0}$ work on both sides of Eq. (\ref{eq:gauss}). To calculate the required correlator, Eq. (\ref{eq:corr2}), we replace $\vec{j} \to \vec{j} + \vec{b}$ in Eq. (\ref{eq:gauss}), where $\vec{b}$ consists of the components $b_i = \sum_j A_j \delta_{ij}$. By letting $\left.\partial_{j_n}\right|_{\vec{j}=0}$ work on both sides of this equation we arrive at
\begin{equation}
\left\la \phi_n e^{\vec{b}^T\cdot \vec{\phi}}\right\ra = \bar{G}_{nl}b_le^{(1/2)\vec{b}^T\cdot\bar{G}^{-1}\vec{b}}.
\end{equation}
Here we sum over double indices. We also used the fact that $\bar{G}^{-1}$ is a symmetric matrix, e.g., $G^{-1}_{nm} = G^{-1}_{mn}$, which is true since it contains contour ordered correlation functions. By letting $\left.\partial^2_{j_n j_m}\right|_{\vec{j}=0}$ work on  Eq. (\ref{eq:gauss}) we get
\begin{equation}
\left\la \phi_n \phi_m e^{\vec{b}^T\cdot \vec{\phi}}\right\ra = \left[ \bar{G}^{-1}_{mn} + \bar{G}^{-1}_{ml}b_l \bar{G}^{-1}_{nl}b_l\right]e^{(1/2)\vec{b}^T\cdot\bar{G}^{-1}\vec{b}}.
\end{equation}
Both equations above are valid when $\sum_j A_j = 0$, otherwise the correlators vanish.


\begin{thebibliography}{99}
\bibitem{Ten952}
E.g., in KCuF${}_3$, see Ref. \onlinecite{Ten95}. For different materials the exchange interaction can be significantly different, see e.g., Ref. \onlinecite{Mot96}, which reports that $J \approx 2\cdot10^3$ K for Sr${}_2$CuO${}_3$ and SrCuO${}_2$.
\end{thebibliography}
\end{document}